\documentclass[11pt,twoside]{article}
\usepackage{asp2006marc}
\usepackage{psfig}
\usepackage{epsf}
\usepackage{graphics}
\usepackage{graphicx}
\usepackage{lscape}
\def\edcomment#1{\iffalse\marginpar{\raggedright\sl#1\/}\else\relax\fi}
\reversemarginpar

\begin{document}
\title{Advanced Technology Large-Aperture Space Telescope (ATLAST): Characterizing Habitable Worlds}

\begin{quote}
M. Postman$^1$, W. Traub$^2$, J. Krist$^2$, K. Stapelfeldt$^2$, R. Brown$^1$,\\
W. Oegerle$^3$, A. Lo$^4$, M. Clampin$^3$, R. Soummer$^1$, J. Wiseman$^3$, and M. Mountain$^1$\\

\footnotesize
{\itshape $^1$Space Telescope Science Institute, Baltimore, MD, USA}\\
{\itshape $^2$Jet Propulsion Laboratory, California Inst. of Technology, Pasadena, CA, USA}\\
{\itshape $^3$NASA/Goddard Space Flight Center, Greenbelt, MD, USA}\\
{\itshape $^4$Northrop Grumman Aerospace Systems, Redondo Beach, CA, USA}\\
\normalsize
\end{quote}

\vskip -15pt
\begin{abstract}
The Advanced Technology Large Aperture Space Telescope (ATLAST) is a set of mission 
concepts for the next generation UV-Optical-Near Infrared space telescope with an 
aperture size of 8 to 16 meters. ATLAST, using an internal coronagraph or an external occulter, can characterize 
the atmosphere and surface of an Earth-sized exoplanet in the Habitable Zone of long-lived stars at 
distances up to $\sim$45 pc, including its rotation rate, climate, and habitability. ATLAST will also allow us to glean 
information on the nature of the dominant surface features, changes in cloud 
cover and climate, and, potentially, seasonal variations in surface vegetation. ATLAST will be able 
to visit up to 200 stars in 5 years, at least three times each, depending on the technique used for 
starlight suppression and the telescope aperture. More frequent visits can be made for interesting systems.
\end{abstract}

\section{Overview of the ATLAST Concept}
We are at the brink of answering two paradigm-changing questions: Do Earth-sized planets 
exist in the Habitable Zones of their host stars? Do any of them harbor life? The tools for answering 
the first question already exist (e.g., Kepler, CoRoT); those that can address the second can be 
developed within the next 10-20 years (Kasting, Traub et al. 2009).  ATLAST is our best option for an extrasolar life-finding 
facility and is consistent with the long-range strategy for space-based initiatives recommended by the 
AAAC Exoplanet Task Force \citep{Lunine2008}. ATLAST is a NASA Astrophysics Strategic Mission Concept for the next 
generation flagship UVOIR space observatory (wavelength coverage: 110 nm -- 2400 nm), designed to answer 
some of the most compelling astronomical questions, including ``Is there life elsewhere in the Galaxy?'' 
The ATLAST team investigated two different observatory architectures: a telescope with an 8-m 
monolithic primary mirror and two variations of a telescope with a large segmented primary 
mirror (9.2-m and 16.8-m). The two architectures span the range in viable technologies: e.g., 
monolithic vs. segmented apertures, Ares V launch vehicle vs. EELV, and passive vs. fully active wavefront 
control. This approach provides several pathways to realize the mission that would be 
narrowed to one as the needed technology development progresses and the availability of 
launch vehicles is clarified. 

While ATLAST requires some technology development, all the observatory concepts take full 
advantage of heritage from previous NASA missions, as well as technologies available 
for missions in development.  The 8 m monolith architecture is similar to the Hubble 
Space Telescope (HST), although the optical design is different.  The 8 m mirror has 
an areal density exceeding that of the HST mirror, providing superb stiffness and thermal inertia.  The 
use of such a massive mirror provides excellent wavefront quality and is made possible 
given a launch vehicle with the capabilities of the proposed Ares V.  The 9.2 m 
and 16.8 m segmented mirror concepts rely heavily on design heritage from the James Webb Space 
Telescope (JWST), in development of lightweight segmented optics, the OTA deployment mechanics, and 
wavefront sensing and control.  The 9.2 m segmented concept can be launched on a Delta IV Heavy 
launch vehicle with a modified 6.5 m fairing; the 16.8 m concept requires an Ares V. The 
non-cryogenic nature of ATLAST makes the construction and testing of the 
observatory much simpler than for JWST.  We have also identified departures from existing NASA 
mission designs to capitalize on newer technologies, minimize complexity, and enable 
the required improvements in performance.

Four significant drivers dictate the need for a large space-based 
telescope if one wishes to conduct a successful search for biosignatures on exoplanets. 
First, and foremost, Earth-mass planets are faint -- an Earth twin at 10 pc, seen at maximum elongation 
around a G-dwarf solar star, will have V $\sim$ 29.8 AB mag. Detecting a 
biosignature, such as the presence of molecular oxygen in the exoplanet's 
atmosphere, will require the ability to obtain direct low-resolution spectroscopy of such 
extremely faint sources. Second, the average projected angular radius of the Habitable Zone (HZ) around 
nearby F,G,K stars is less than 100 milli-arcseconds (mas). 
One thus needs an imaging system capable of angular resolutions of $\sim$10 to 25 mas 
to adequately sample the HZ and isolate the exoplanet point source in the presence of an exo-zodiacal 
background. Third, direct detection of an Earth-sized planet in the HZ requires high contrast 
imaging, typically requiring starlight suppression factors of 10$^{-9}$ to 10$^{-10}$.  
Several techniques \citep{Levine2009} are, in principle, capable of delivering such high contrast levels but 
all require levels of wavefront stability not possible with ground-based telescopes because 
the timescale of wavefront variations induced by the Earth's atmosphere is shorter than the time to measure the 
wavefront error to the required precision. A space-based 
platform is required to achieve the wavefront stability that is needed for such high contrast imaging. 
{\footnote {Such stability is achievable regardless of the space telescope's 
aperture (within reasonable limits) -- so this driver is primarily for the space locale of 
the telescope rather than its aperture.}} 
Lastly, biosignature-bearing planets may 
well be rare, requiring one to search tens or even several hundred stars to find even a 
handful with compelling signs of life. The number of stars for which one can 
obtain an exoplanet's spectrum at a given SNR and in less than a given exposure time scales approximately 
as D$^3$, where D is the telescope aperture diameter.  This is demonstrated in Figure 1 
where we have averaged over different simulations done using various starlight suppression options 
(internal coronagraphs of various kinds as well as an external occulter). 
To estimate the number of potentially habitable worlds detected, one must multiply the 
numbers in Figure 1 by the fraction of the stars that have an exoplanet with detectable biosignatures 
in their HZ ($\eta_{\oplus}$). The value of $\eta_{\oplus}$ is currently not constrained 
but it is not likely to be close to unity. One must conclude that to maximize the chance for 
a successful search for life in the solar neighborhood requires a space telescope with 
an aperture size of at least 8 meters.

\setcounter{figure}{0}
\begin{figure}
\hskip 10pt
\includegraphics[height=1.6in]{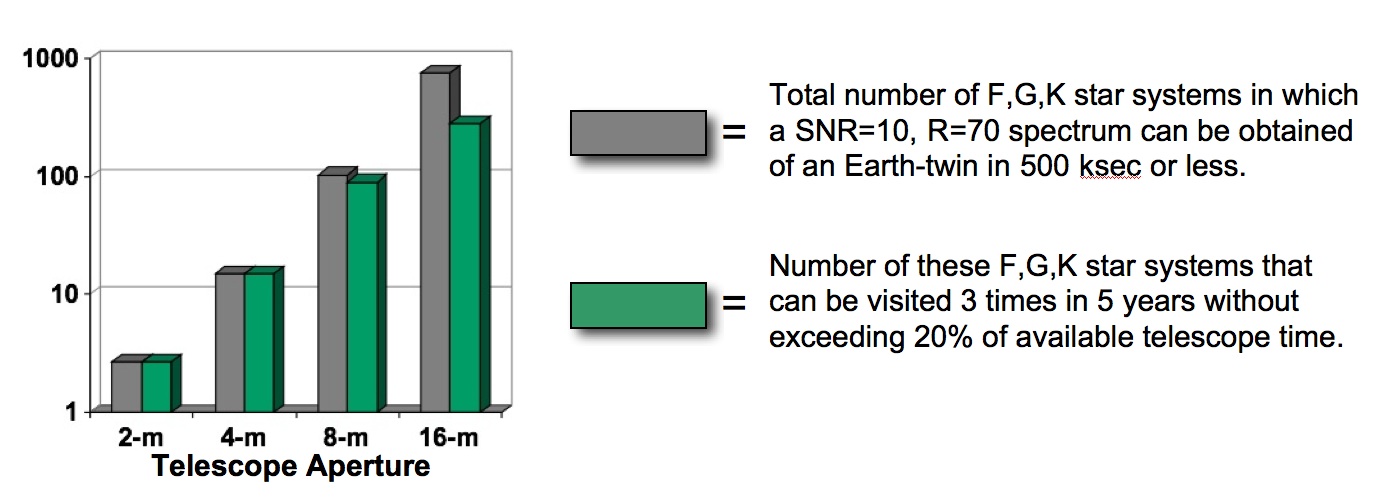}
\caption{\footnotesize The number of F,G,K stars as a function of telescope
aperture where an R=70 SNR=10 spectrum could be obtained of an Earth-twin
in the HZ in less than 500 ksec. These counts assume every star has an Earth twin.}
\end{figure}

All ATLAST concepts require many of the same key technologies. 
We believe these designs compose a robust set of options for achieving the next generation of 
UVOIR space observatory in the 2020 era. 
There are several fundamental features common to all our designs. 
All ATLAST concepts are designed to operate in a halo orbit at the Sun-Earth L2 point. 
The optical designs are diffraction limited at 500 nm (36 nm RMS WFE) and the optical telescope 
assembly (OTA) operates near room temperature (280 K -- 290 K). All OTAs employ two, simultaneously usable 
foci:  a three-mirror anastigmat channel for multiple, wide field of view instruments, and a 
Cassegrain channel for high-throughput UV instruments and instruments for imaging and 
spectroscopy of exoplanets (all designs have an RMS WFE of $<5$ nm at $<2$ arcsec radial offset from Cass optical axis). 
The ATLAST concept study is available on line and on the astro-ph archives \citep{ATLAST1,ATLAST2}.

\section{Simulated Data and Exoplanet Characterization Performance}

Estimates of the SNR of habitability and biosignature features in an Earth-twin 
spectrum, achievable with ATLAST, are shown in Table 1. For these calculations we use a fully 
validated model of the Earth's spectrum \citep{Marais2002,Woolf2002}, in combination with the observed visible reflection 
spectrum of the present Earth.  We assume that the exoplanet is at maximum elongation and that 
the planet is observed for a length of time sufficient to achieve an SNR of 10, at a spectral resolution 
R = 70, in the red continuum. 
The vegetation signal is the enhanced albedo of the Earth from land plants, for wavelengths longer 
than ~720 nm \citep{Seager2005}, with a modest SNR.  
Column 3 gives the width of the spectral feature. All of these SNR values can easily be improved with 
re-visits. In addition, ATLAST will allow us to glean substantial information about an exo-Earth from 
temporal variations in its features. Such variations inform us about the nature of the dominant 
surface features, changes in climate, changes in cloud cover, and, potentially, seasonal 
variations in surface vegetation \citep{Ford2003}. Constraints on variability require multiple visits to each target. 
The 8-m ATLAST (with internal coronagraph) will be able to observe $\sim$100 different star 
systems 3 times each in a 5-year interval and not exceed 20\% of the total observing time 
available to the community. The 16-m version (with internal coronagraph) could visit up to 
$\sim$250 stars three times each in a five-year period. The 8-m or 16-m ATLAST (with a single external 
occulter) can observe $\sim$85 stars 3 times each in a 5-year period, limited by the transit times of the occulter. 
Use of two occulters would remove this limitation.
\vskip -5pt
\begin{table}[!h]
\caption{Habitability and Bio-Signature Characteristics}
\smallskip
\begin{center}
{\small 
\begin{tabular}{l|c|c|c|l}
\tableline\tableline
\noalign{\smallskip}
Feature & $\lambda$ (nm) & $\Delta\lambda$ (nm) & SNR & Significance \\
\noalign{\smallskip}
\tableline
\noalign{\smallskip}
Reference Continuum & 750 & 11 & 10 & \\
Rayleigh Scattering & $<$500 & 100 & 4 & Protective atmosphere \\
Ozone (O$_3$) & 580 & 100 & 5 & Source is O$_2$; UV Shield \\
Oxygen (O$_2$) & 760 & 11 & 5 & Plants emit; Animals inhale \\
Cloud/Surface reflection & 750 & 100 & 30 & Rotation indicator \\
Land Plant reflection & 770 & 100 & 2 & Vegetated land area \\
Water vapor (H$_2$O) & 940 & 60 & 16 & Needed for life \\
\noalign{\smallskip}
\tableline
\end{tabular}
}
\end{center}
\end{table}

\begin{figure}
\hskip -20pt
\includegraphics[width=5.5in]{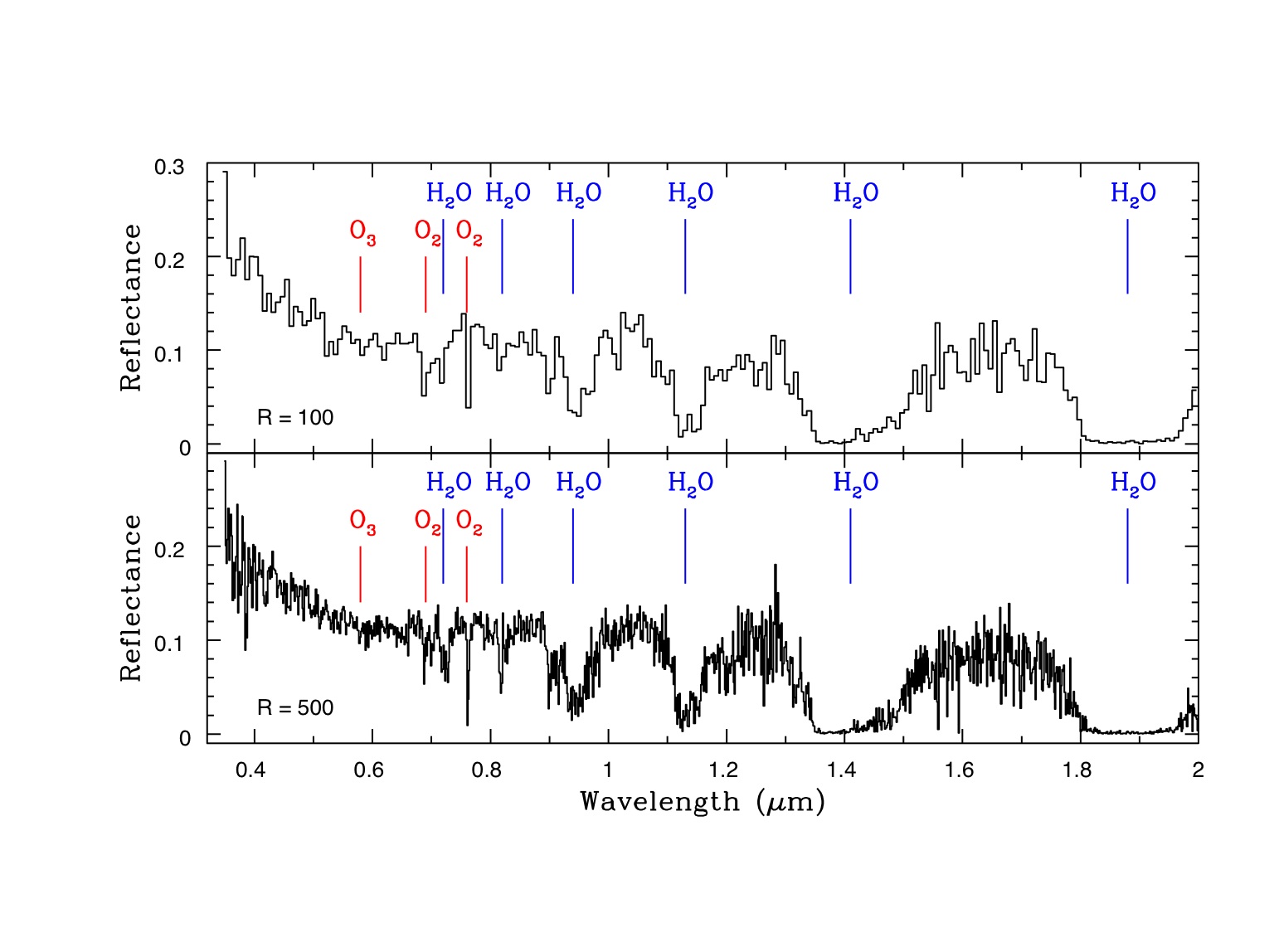}
\caption{\footnotesize Simulated ATLAST spectrum of an Earth-twin at 10 pc, shown at R=100 in the upper 
plot and R=500 in the lower plot. The SNR=10 at 750 nm in both cases. Key O$_2$, O$_3$, 
and H$_2$O features are shown. The increased reflectance at the blue end of 
the spectrum is due to Rayleigh scattering. An 8-m ATLAST obtains 
the R=100 and R=500 spectra in 46 ksec and 500 ksec, respectively.}
\end{figure}
\vskip -12pt
Figure 2 shows two simulated ATLAST spectra for an Earth-twin at 10 pc, one at R=100 and one at R=500, taken 
with sufficient exposure to reach SNR=10 at 750 nm in the continuum. A 3-zodi background was 
used (local plus exosolar). The R=100 exposure times are 46 ksec and 8 ksec, respectively, for an 8-m 
and 16-m space telescope. The corresponding exposure times for the R=500 spectrum are 500 ksec and 56 ksec, 
respectively, for the 8-m and 16-m telescopes. The reflected flux from an Earth-like rocky planet 
increases as M$^{2/3}$, where M is the exoplanet mass. Hence, the exposure times for a 5M$_{\oplus}$ super-Earth 
would be $\sim$3 times shorter. At both resolutions, the O$_2$ features 
at 680 nm and 760 nm are detected, as are the H$_2$O features at 720, 820, 940, 1130, 
1410, and 1880 nm. Rayleigh scattering is detected as an increase in reflectivity bluewards of 550 nm. 
The higher spectral resolutions enabled by large-aperture 
space telescopes enable the detection of molecular oxygen in exoplanets with lower abundances 
than those on Earth and provide constraints on the kinematics and thermal structure of the 
atmosphere that are not accessible at lower resolution.

\begin{figure}
\includegraphics[width=5in]{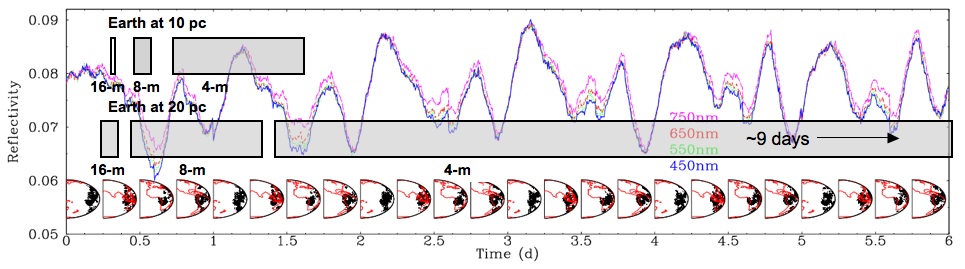}
\caption{\footnotesize Model light curves of Earth for a six day interval (Ford et al. 2003). Superposed are grey bands whose length 
indicates the time required (in days) to 
achieve SNR=20 broadband photometry of an Earth twin at 10 pc (top bands) and 20 pc (bottom bands) for space 
telescopes with apertures of 4, 8, and 16 meters. The height of the grey bands corresponds 
to the $\pm$5\% photometric error.}
\end{figure}

For a 16-m class space telescope, time-resolved spectroscopy over intervals of a few hours may reveal 
surface composition variations, if the planet is not cloud dominated, as the exoplanet rotates. However, even broadband 
photometry can be used to detect short-term variations in albedo that can determine the rotation period 
and constrain the amount of cloud cover. Ford et al. (2003) generated model light curves 
for the Earth over 6 consecutive days using data from real satellite observations. Photometric variations 
of 20 -- 30\% on timescales of 6 hours were typical in the B,V,R,I passbands. Their models are shown
in Figure 3. On top of these light curves we show grey bands whose height represents the $\pm$5\% uncertainty 
for SNR=20 broadband (R=4) photometry and whose length represents the time it would take to perform 
such an observation with a 4-m, 8-m, and 16-m space telescope. We show the results of such 
calculations for an Earth twin at 10 pc and 20 pc. The 4-m has marginal capability to study such photometric 
variations at 10 pc but telescopes with apertures of 8-m or larger would be able to perform the 
measurements well as the integration time is less than the typical period between significant albedo changes. 
At 20 pc, even an 8-m telescope reaches its limits but a 16-m telescope is 
still able to acquire the needed accuracy in photometry in less than 4 hours.

Transit spectroscopy with ATLAST will  permit characterization of super-Earth mass exoplanets.
Figure 4 shows two simulated ATLAST transit spectra for planets around a 6th magnitude M2 star where the orbital period in the 
HZ is $\sim$20 -- 30 days. Such observations are time consuming but do not require the use of a coronagraph or 
occulter.

\begin{figure}
\hskip 10pt
\includegraphics[height=2in]{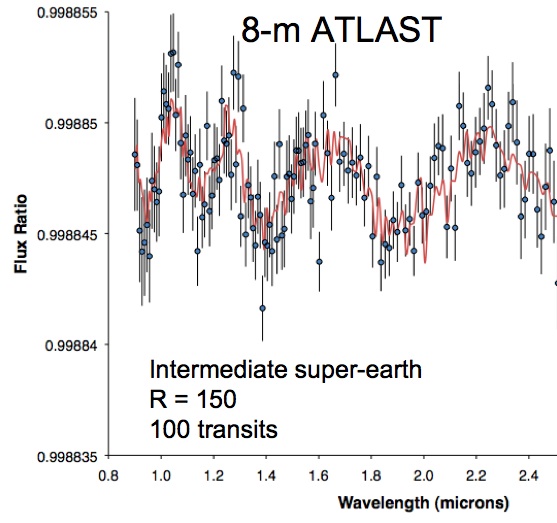}
\hskip 20pt
\includegraphics[height=2in]{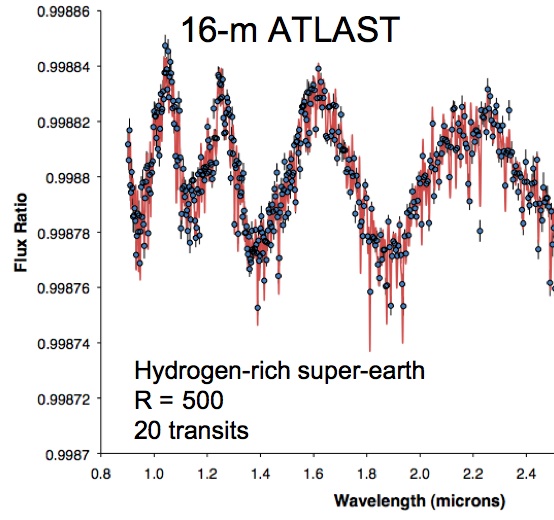}
\caption{\footnotesize Simulated ATLAST transit spectra for two super-Earth exoplanets around a K=6 mag M2 star. Instrumental
effects are modeled assuming JWST NIRSpec performance. The transit period is $\sim$22 days. The broad features are water absorption bands.}
\end{figure}

In summary, with ATLAST, we will be able to determine if HZ exoplanets are indeed habitable, 
and if they show signs of life as evidenced by the presence of oxygen, water, and ozone. 
ATLAST also will provide useful information on the column abundance of the atmosphere, the presence of 
continents and oceans, the rotation period, and the degree of daily large-scale weather variations.

\end{document}